\documentclass[preprint,aps,floats,graphicx,epsf]{revtex4}
\def\be{\begin{equation}}
\def\ee{\end{equation}}
\def\bea{\begin{eqnarray}}
\def\eea{\end{eqnarray}}
\usepackage{graphicx}% Include figure files% allows one to include
\begin{document}
%\preprint{APS/123-QED}

\title{\bf Turbulent-Like Behavior of Seismic Time Series}

\author{P. Manshour,$^1$ S. Saberi,$^1$ Muhammad Sahimi,$^2$ \\ J. Peinke,$^3$
Amalio F. Pacheco,$^4$, M. Reza Rahimi Tabar,$^{1,3,5}$ \\
{$^1$Department of Physics, Sharif University of Technology, Tehran
11155-9161, Iran\\
$^2$Mork Family Department of Chemical Engineering \& Materials Science,
University of Southern California, Los Angeles, California 90089-1211, USA\\
$^3$Carl von Ossietzky University, Institute of Physics, D-26111
Oldenburg, Germany\\
$^4$Department of Theoretical Physics, University of Zaragoza,\\
Pedro Cerbuna 12, 50009 Zaragoza, Spain\\
$^5$CNRS UMR 6202, Observatoire de la C$\hat o$te d'Azur, BP 4229,
06304 Nice Cedex 4, France}}

%\maketitle

\begin{abstract}

\bigskip

We report on a novel stochastic analysis of seismic time series for
the Earth's vertical velocity, by using methods originally developed
for complex hierarchical systems, and in particular for turbulent flows.
Analysis of the fluctuations of the detrended increments of the
series reveals a pronounced change of the shapes of the probability
density functions (PDF) of the series' increments. Before and close to
an earthquake the shape of the PDF and the long-range correlation in
the increments both manifest significant changes. For a moderate or
large-size earthquake the typical time at which the PDF undergoes the
transition from a Gaussian to a non-Gaussian is about 5-10 hours.
Thus, the transition represents a new precursor for detecting such
earthquakes.

\bigskip

\noindent PACS number(s): 05.45.Tp 64.60.Ht 89.75.Da 91.30.Px

\end{abstract}
\maketitle

\vskip 0.5cm

A grand challenge in geophysics, and the science of analyzing seismic
activities, is developing methods for predicting when earthquakes may occur.
Any accurate information about an impending large earthquake may reduce
the damage to the infrastructures and the number of casualties. Although there
are many known precursors, experience over the last several decades indicates
that reliable and quantitative methods for analyzing seismic data are still
lacking [1].

A large number of stations around the world currently perform precise
measurements of seismic time series. Several concepts and ideas
have been advanced over the past few decades in order to explain
important aspects of such seismic time series and what they imply for
earthquakes, ranging from the Gutenberg-Richter law [2] for the number
of earthquakes with a magnitude greater than a given value $M$, to the
Omori law [3] for the distribution of the aftershocks, the concept
of self-organized criticality [4], and the percolation model of the epicenters'
spatial distribution [5]. At the same time, there has also been much
interest in investigating the precursors to, and the predictability of,
extreme increments in time series [6] associated with disparate phenomena,
ranging from earthquakes [7,8], to epileptic seizures [9], and stock market
crashes [10-12].

In this Letter we provide compelling evidence for the existence of a
novel transition in the probability density function (PDF) of the
detrended increments of the stochastic fluctuations of the Earth's
vertical velocity $V_z$, collected by broad-band stations (resolution
100 Hz). As an important new result, we demonstrate that there
is a strong transition from a Gaussian to a non-Gaussian behavior
of the increments' PDF as an earthquake is approached. We characterize
the non-Gaussian nature of the PDF of the fluctuations in the increments
of $V_z$, and the time-dependence of the PDF of the background fluctuations
far from earthquakes.

The results presented in this Letter are based on the detailed analysis
of the data obtained from Spain's and California's broad-band networks for
three earthquakes of large and intermediate magnitudes: the
May 21, 2003, $M=7.1$ event in Oran-Argel, detected in Ibiza (Balearic
Islands); the 2004, $M=6.1$ event in Alhucemas, and the $M=5.4$ earthquake
that occurred in California on April 30, 2008. Due to our recent discovery
of localization of elastic waves in rock, both experimentally [13] and
theoretically [14], we choose the distance of the detectors from
the epicenters to be three times less than 300 km and one time 400 km. We have also analyzed many
other earthquakes around the world, the results of which will
be described briefly.

The data are first detrended over different time scales in order to remove
the possible trends in the time series $x(t)\equiv V_z(t)$. To do so, the
time series is divided into semi-overlapping subintervals $[1+s(k-1),s(k+1)]$
of length $2s$ and labeled by the index $k\geq 1$. Next, we fit $x(t)$ to a
third-order polynomial [15-17] to detrend the original series in the
corresponding time window. The detrended increments on scale $s$ are defined by
$Z_s(t)=x^*(t+s)-x^*(t)$, where $t\in[1+s(k-1), sk]$, with $x^*(t)$ being the
detrended series, i.e., the deviation of $x(t)$ from its fitted value.

We then develop a new approach, originally proposed for fully-developed
turbulence [18-21], in order to describe the cascading process that
determines how the fluctuations in the series evolve, as one passes from
the coarse to fine scales. For a fixed $t$, the fluctuations at scales $s$
and $\lambda s$ are related through the cascading rule,
\begin{equation}
Z_{\lambda s}(t)=W_\lambda Z_s(t),\hskip 1cm\forall s,\;\lambda>0\;,
\end{equation}
where $\ln(W_\lambda)$ is a random variable. Iterating Eq. (1) forces
implicitly the random variable $W_\lambda$ to follow a log
infinitely-divisible law [22]. One of the simplest candidates for
such processes is represented [17] by, $Z_s(t)=\zeta_s(t)\exp[\omega_s(t)]$,
where $\zeta_s$ and $\omega_s(t)$ are independent Gaussian variables with
zero mean and variances $\sigma_\zeta^2$ and $\sigma_\omega^2$.
The PDF of $Z_s(t)$ has fat tails that depend on the variance of
$\omega_s$, and is expressed by [21],
\begin{equation}
P_s(Z_s)=\int F_s\left(\frac{Z_s}{\sigma}\right) \frac{1}{\sigma}
G_s(\ln\sigma)d\ln\sigma\;,
\end{equation}
where $F_s$ and $G_s$ are both Gaussian with zero mean and variances
$\sigma_s^2$ and $\lambda_s^2$, respectively, e.g., $G_s(\ln\sigma)=
1/(\sqrt{2\pi}\lambda_s)\exp(-\ln^2\sigma/2\lambda^2_s)$. In this
case, $P_s(Z_s)$ is expressed by Eq. (2) and converges to a Gaussian
distribution as $\lambda_s^2\to 0$. Although Eq. (2) is equivalent
to that for a log-normal cascade model, originally introduced to
study fully-developed turbulence [23,24], it also describes
approximately the non-Gaussian PDFs observed in a broad range of
other phenomena and systems, such as foreign exchange markets
[17,24,25] and heartbeat interval fluctuations [17,18] (see also
[26-28]).

To carry out a quantitative analysis of the seismic times series, we focus
on two aspects, namely, deviations of the PDF of the detrended increments
from a Gaussian distribution, and the dependence of correlations in the
increments on the scale parameter $s$. We begin with the time series for the
largest ($M=7.1$) earthquake for two distinct time intervals: (i) data set (I)
representing the background fluctuations far from the time of the earthquake,
and (ii) data set (II) close (less than 5 hours) to the earthquake.

To fit the increments' PDF to Eq. (2), we estimate the variance
$\lambda^2_s(s)$, using the least-squares method, with the error bars estimated
by the goodness of the fit method. Deviation of $\lambda^2_s(s)$ from zero is
a possible indicator of non-Gaussian statistics. As shown in Fig. 1, we find
an accurate parametrization of the PDFs by $\lambda^2_s(s)$ for both data sets.
Moreover, the PDF of $Z_s$ for the data set (I) becomes essentially Gaussian
as $s$ increases to 800 ms, whereas it deviates from the Gaussian distribution
for the data set (II). The time scale $s=800$ ms for $\lambda_s^2$ within a
moving window was estimated by the plot of $\lambda_s^2$ vs $s$ for the
data set (I) (background fluctuations), and selecting $s$ such that
$\lambda_s\to 0$ (see also below).

The scale-dependence of the parameter $\lambda_s^2$ of the PDF is shown in
Fig. 2. For the data set (I) of the $M=7.1$ earthquake, shown in Fig. 2(a), and
times $200\;{\rm ms}<s<500\;{\rm ms}$, a logarithmic decay, $\lambda^2_s\propto
\log s$, is obtained. For the data set (II), the logarithmic regime extends to
$300\;{\rm ms}<s<2000\;{\rm ms}$. Figure 2(b) presents similar behavior for the
$M=6.1$ earthquake. We note that, for the data set (II) of the $M=6.1$
earthquake, there is a crossover time at which $\lambda_s^2$ changes from a
$\sim\log(s)$ behavior to having a finite value, $\simeq 0.3$.

The importance of the results shown in Fig. 2 is that, they indicate that
the increments' PDFs for $s>2000$ ms and $s>1500$ ms are almost Gaussian
($\lambda_s^2\to 0$) for the $M=7.1$ and $M=6.1$ earthquakes [for the data set
(II)], respectively. Transforming the time scales to length scales via the
velocity of the elastic waves in Earth, $\sim 5000$ m/sec, the corresponding
length scales are about 10 km and 7.5 km, for the same earthquakes,
respectively, implying that larger earthquakes have larger characteristic
length scales, and that for the $M=6.1$ event, there is a smaller active part
in the fault.

We note that as one moves down the cascade process from the large to small
scales, one expects the statistics to increasingly deviate from Gaussianity
(see above and Fig. 2), in order to derive Eq. (2). Note that, a non-Gaussian
PDF with fat tails on small scales indicates an increased probability of the
occurrence of short-time {\it extreme} seismic fluctuations.

From the point of view of the increments' PDF, the non-Gaussian noise with
uncorrelated $\omega_s$ in the process, $Z_s(t)=\zeta_s(t)\exp[\omega_s(t)]$,
and a multifractal formulation are indistinguishable, because their
one-point statistics at any given scale may be identical. Thus, to
understand the origin of the non-Gaussian fluctuations, we explore the
correlation properties of $\omega_s$. To do so, we use an alternative method
for studying the correlation functions of the local ``energy'' fluctuations
[20]. We define the magnitude of local variance over a scale $s$ by,
$\sigma_s^2(t)_s^{-1} \sum^{n_s/2}_{k=-n_s/2}Z_s(t+k\Delta t)^2$, and,
$\bar{\omega}_s(i)=\frac{1}{2} \log\sigma_s^2(i)$, respectively. Here, $\Delta
t$ is the sampling interval and, $n_s\equiv s/\Delta t$. The magnitude of the
correlation function of $\bar{\omega_s}$ is then defined by
\begin{equation}
C^{(s)}(\tau)=\langle[\bar{\omega}_s(t)-\langle\bar{\omega}_s
\rangle]
[\bar{\omega}_s(t+\tau)-\langle\bar{\omega}_s\rangle]\rangle\;,
\end{equation}
where $\langle\cdot\rangle$ indicates a statistical average. Figure 3 shows
the results for the two data sets. The correlation function decays sharply for
the data set (I) - far from the earthquakes - whereas it is of long-range type
for the set (II) - close to the earthquakes.

We emphasize that for the data set (II), the PDF deviates from being Gaussian
even for $s > 800$ ms. Although one might argue that the deviations might be
due to an underlying L\'evy statistics, this possibility is ruled out due to
the deduced hierarchical structures that imply that the increments for
different scales are {\it not} independent; see Fig 3.

We now show that the analysis may be used as a new precursor for detecting
an impending earthquake. A window containing one hour of data is selected and
moved with $\Delta t=15$ minutes to determine the temporal dependence of
$\lambda_s^2$. Guided by Fig. 2, the local temporal variations of $\lambda_s^2$
for $s=800$ ms are investigated. According to Fig. 2, for $s\simeq 800$ ms, the
difference between the values of $\lambda^2$ is large enough for the background
data and the data set near the earthquakes. Hence, such a time scale may be
used as the characteristic time for the dynamics of the non-Gaussian indicator
$\lambda_s^2$. Figures 4(a) and 4(b) display a well-pronounced, systematic
increase in $\lambda_s^2$ as the earthquakes are approached. Taking into
account the estimated error of $\lambda_s^2$ for the background fluctuations in
Figs. 4, we see that about 7 and 5 hours before the earthquakes values of
$\lambda_s^2$ are larger, by more than two standard deviations, than those for
the background.

We note that in defining $\lambda_s$, one supposes that the PDF of the
increments is log-normal. This may induce errors due to the fact that, one
must fit the PDF with a predefined functional form. An unbiased quantity that
measures the intermittency and deviation from Gaussian is the flatness, which
provides an unbiased estimator of deviations from Gaussianity, without having
to adopt {\it a priori} any functional form for the PDF. In Figs. 4(c) and 4(d)
we present the flatness of the time series in the same windows (see above) for
time scale $s \simeq 800$ ms. They show that, consistent with $\lambda_s^2$,
the flatness also yields a clear alert for an impending earthquake.

Due to the localization of elastic waves in Earth, stations that are far from
an earthquake epicenter cannot provide any clue to the transition in the shape
of the increments' PDF. We checked this point for several earthquakes. Shown in
Fig. 5 are the results for the $M=5.4$ California earthquake, occurred
at (40.837 N, 123.499 W). The distance from the epicenter of the station that
does provide an alert of about 3 hours for the earthquake is about $\simeq 128$
km, while the second station that does not yield any alert is about $\simeq
400$ km from the epicenter.

We also analyzed several other earthquakes of various magnitudes. We found that
for earthquakes with $M\leq 5$ the increase in $\lambda_s^2$ is not large, even
for the data that are collected in stations about 100 km from the epicenters.
Moreover, we also analyzed seismic data for large earthquakes in Pakistan and
Iran, and found that they exhibit the same types of trends and results, as
those presented above, for the time variation of $\lambda_s^2$ close to the
earthquakes. For example, for the $M=7.6$ earthquake that occurred on August
10, 2005, in Pakistan, the transition in the value of $\lambda_s^2$ occurred
about 10 hours before the earthquake, while for the $M=6.3$ earthquake that
occurred in northern Iran on May 28, 2004, the transition happened about 4
hours before the earthquake.

In summary, the temporal dependence of the fat tails of the PDF of the
increments of the vertical velocity $V_z(t)$ of Earth exhibits a gradual,
systematic increase in the probability of the appearance of large values on
approaching a large or moderate earthquake, which is interpreted as an alert
for the earthquake. To estimate the alert time one must, (i) utilize the time
series $V_z(t)$, collected at broad-band stations near the epicenters.
Due to localization of elastic waves in rock [15], data from far away stations
cannot provide any clue to the transition in the shape of the increments' PDF.
The station's distance (300 km) is {\it not} universal and depends on the
geology, but it is of the correct order of magnitude. (ii) The time scale $s$
for moving $\lambda_s^2$ (here, 800 ms) is estimated by its plot vs $s$ for
data set (I), and a choice of $s$ such that $\lambda^2\to 0$. On this scale
the difference between $\lambda_s^2$ for the data sets (I) and (II) will be
large enough to obtain a meaningful alert for the earthquake. (iii) One must
also estimate $\lambda_s^2$ or the flatness in some windows (here, 1 hour
windows) and move it over the time series, in order to observe its variations
with the time.

We thank R. Friedrich, U. Frisch, H. Kantz, and K. R. Sreenivasan for critical
reading of the manuscript, as well as J. G. Jim$\acute{e}$nez, T. Matsumoto,
M. Mokhtari and S. M. Movahed for useful discussions. M.R.R.T. would like to
thank the Alexander von Humboldt Foundation, the Associate program of the
Abdus Salam International Center for Theoretical Physics, and the  {\it
Knowledge Archive} funding for financial support. The data for $V_z(t)$
were provided by the USGS and the National Geographic Institute of Spain.

\newpage

\newcounter{bean}
\begin{list}%
{[\arabic{bean}]}{\usecounter{bean}\setlength{\rightmargin}{\leftmargin}}

\item {\it Modelling Critical and Catastrophic Phenomena in Geoscience: A
Statistical Physics Approach}, edited by P. Bhattacharyya and B. K.
Chakrabarti, Lecture Notes in Physics {\bf 705} (Springer, Berlin,
2006).

\item B. Gutenberg and R. F. Richter, {\it Seismicity of the Earth} (Hafner
Publishing, New York, 1965).

\item T. Utsu, Y. Ogata, and R. S. Matsu'ura, J. Phys. Earth {\bf 43}, 1
(1995).

\item P. Bak and C. Tang, J. Geophys. Res. {\bf 94}, 15635 (1989).

\item M. Sahimi, M. C. Robertoson, and C. G. Sammis, Phys. Rev. Lett. {\bf 70},
2186 (1993); H. Nakanishi {\it et al.}, Phys. I {\bf 3}, 733 (1993), M. C.
Robertson {\it et al.}, J. Geophys. Res. B {\bf 100}, 609 (1995).

\item S. Hallerberg, E. G. Altmann, D. Holstein, and H. Kantz, Phys. Rev. E {\bf 75}, 016706
(2007); S. Hallerberg and H. Kantz, {\it ibid.} {\bf 77}, 011108 (2008).

\item D. D. Jackson, Proc. Natl. Acad. Sci. USA {\bf 93}, 3772 (1996).

\item M. R. Rahimi Tabar, M. Sahimi, K. Kaviani {\it et al.}, in Ref. [1], p.
281.

\item F. Mormann {\it et al.}, Clin. Neurophysiol. {\bf 116}, 569 (2005).

\item A. Johansen and D. Sornette, Eur. Phys. J. B {\bf 1}, 141 (1998).

\item N. Vandewalle {\it et al.}, Eur. Phys. J. B {\bf 4}, 139 (1998).

\item D. Sornette, Proc. Natl. Acad. Sci. USA {\bf 99}, 2522 (2002).

\item  E. Larose, L. Margerin, B. A. vanTiggelen, M. Campillo, Phys. Rev. Lett. {\bf 93}, 048501 (2004); N. M.
Shapiro {\it et al.}, Science {\bf 307} (2005).

\item F. Shahbazi, A. Bahraminasab, S. M. Vaez Allaei, M. Sahimi, and M. R.
Tabar, Phys. Rev. Lett. {\bf 94}, 165505 (2005); A. Bahraminasab,
S. M. Allaei, F. Shahbazi, M. Sahimi, M. D. Niry, M. R. Tabar,
 Phys. Rev. B {\bf 75}, 064301 (2007); R. Sepehrinia, 
A. Bahraminasab, M. Sahimi, M. R. Tabar,
{\it ibid.} {\bf 77}, 014203 (2008); S. M. Vaez Allaei, M. Sahimi, and M. R.
Rahimi Tabar, J. Stat. Mech. (2008) P03016.

\item K. Kiyono {\it et al.}, Phys. Rev. Lett. {\bf 93}, 178103 (2004).

\item K. Kiyono, Z. R Struzik, N. Aoyagi, F. Togo, Y. Yamamoto,
 Phys. Rev. Lett. {\bf 95}, 058101 (2005).

\item K. Kiyono, Z. R. Struzik, and Y. Yamamoto, Phys. Rev. Lett. {\bf 96},
068701 (2006); Phys. Rev. E {\bf 76}, 041113 (2007); G. R. Jafari,
{\it et al.}, Int. J. Mod. Phys. C {\bf 18}, 1689 (2007).

\item U. Frisch and D. Sornette, J. Phys. I (France) {\bf 7}, 1155 (1997).

\item J. F. Muzy, J. Delour, and E. Bacry, Eur. Phys. J. B {\bf 17} 537 (2000).

\item A. Arneodo, E. Bacry, S. Manneville, J. F. Muzy, Phys. Rev. Lett. {\bf 80}, 708 (1998).

\item B. Castaing, Y. Gagne, and E. J. Hopfinger, Physica D {\bf 46}, 177
(1990).

\item B. Dubrulle, Phys. Rev. Lett. {\bf 73} 959 (1994); Z.-S. She and E. C.
Waymire, {\it ibid.} {\bf 74} 262 (1995).

\item B. Chabaud {\it et al.}, Phys. Rev. Lett. {\bf 73}, 3227 (1994).

\item H. E. Stanley and V. Plerou, Quant. Fin. {\bf 1}, 563 (2001).

\item S. Ghashghaie, W. Breymann, J. Peinke, P. Talkner, and Y. Dodge, Nature
{\bf 381}, 767 (1996).

\item P. A. Varotsos, N. V. Sarlis, H. K. Tanaka, E. S. Skordas, Phys. Rev. E {\bf 72}, 041103 (2005).

\item M. De Menecha and A. L. Stella, Physica A {\bf 309}, 289 (2002).

\item F. Caruso, A. Pluchino, V. Latora, S. Vinciguerra, A. Rapisarda, Phys. Rev. E {\bf 75}, 055101(R) (2007).

\end{list}%

\newpage

\begin{figure*}[htb]
\centerline{\includegraphics[scale=0.7,draft=false]{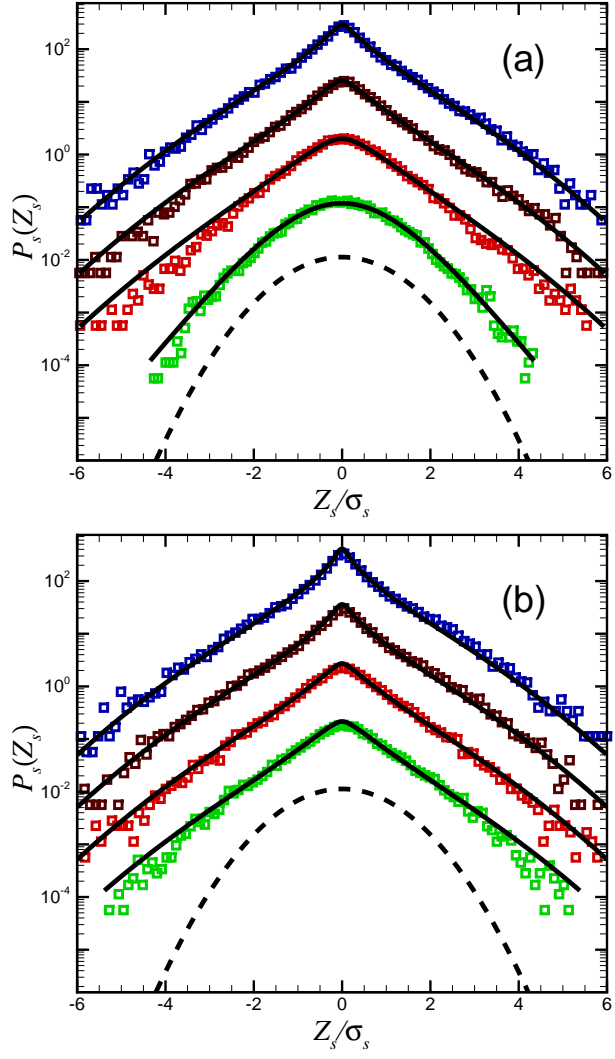}}
\caption{Continuous deformation of the increments' PDFs in log-linear scale,
for the $M=7.1$ earthquake across scales for, from top to bottom, $s=200,\;400,
\;600,$ and 800 ms, and (a) far from, and (b) close to, the earthquake. Solid
curves are the PDFs based on Eq. (2), while dashed curves are the Gaussian
PDF.}
\end{figure*}

\newpage

\begin{figure*}[htb]
\centerline{\includegraphics[scale=0.7,draft=false]{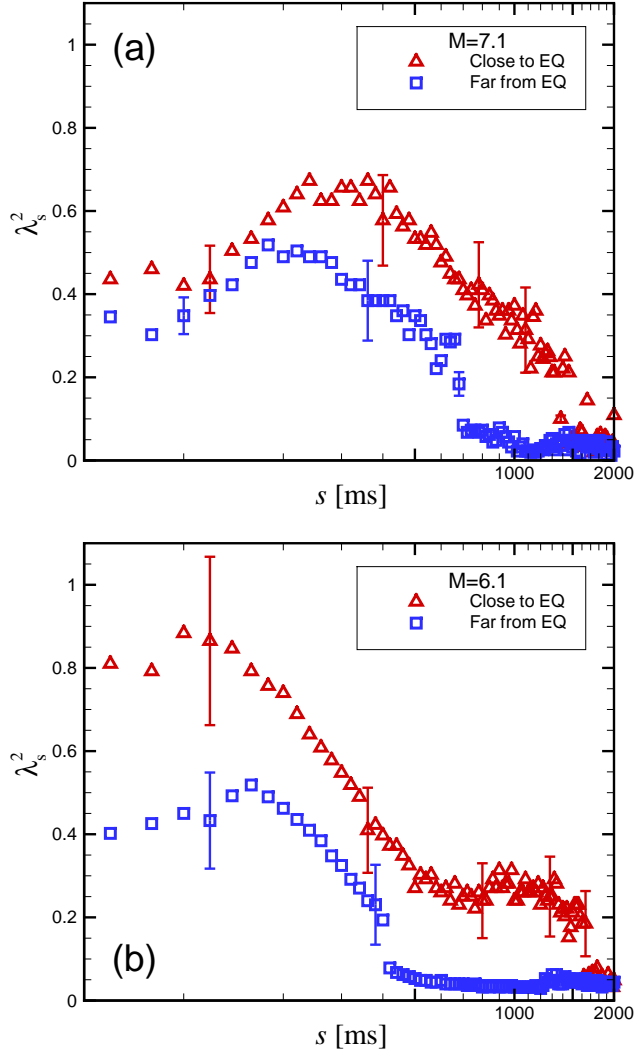}}
\caption{Scale-dependence of $\lambda_s^2$ vs $\log s$. (a) The $M=7.1$ event,
both far from [data set (I)] and close to [data set (II)] the earthquake. For
the data set (I) and $s>700$ ms, $\lambda_s^2\to 0$, implying that the
increments' PDF is Gaussian, whereas for the data set (II) $\lambda_s^2$
deviates strongly from zero for $700\;{\rm ms}<s<1500$ ms. (b) Same as in (a),
but for the $M=6.1$ earthquake. When $\lambda_s\to 0$ the error bars are very
small, about the same size as the symbols.}
\end{figure*}

\newpage

\begin{figure*}[htb]
\centerline{\includegraphics[scale=0.7,draft=false]{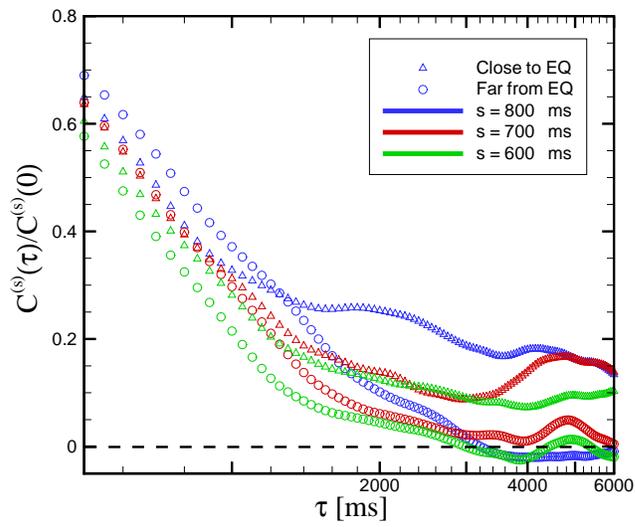}}
\caption{The correlation function $C^{(s)}(\tau)$ for the data set (I) far
from, and the set (II) close to, the $M=7.1$ earthquake.}
\end{figure*}

\newpage

\begin{figure*}[htb]
\centerline{\includegraphics[scale=0.7,draft=false]{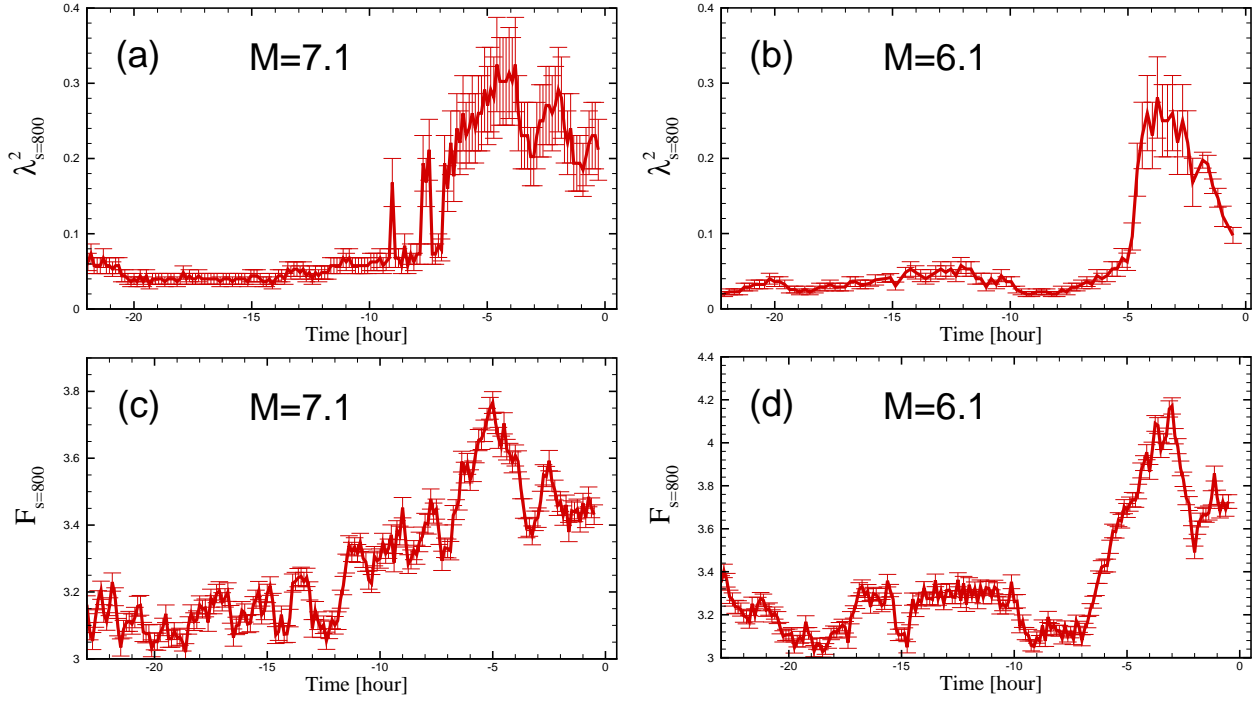}}
\caption{The local temporal dependance of $\lambda^2_s$ and the flatness
for $s=800$ ms, over a one-hour period, for the $M=7.1$ and $M=6.1$ events,
indicating a gradual, systematic increase on approaching the earthquakes.}
\end{figure*}

\newpage

\begin{figure*}[htb]
\centerline{\includegraphics[scale=0.7,draft=false]{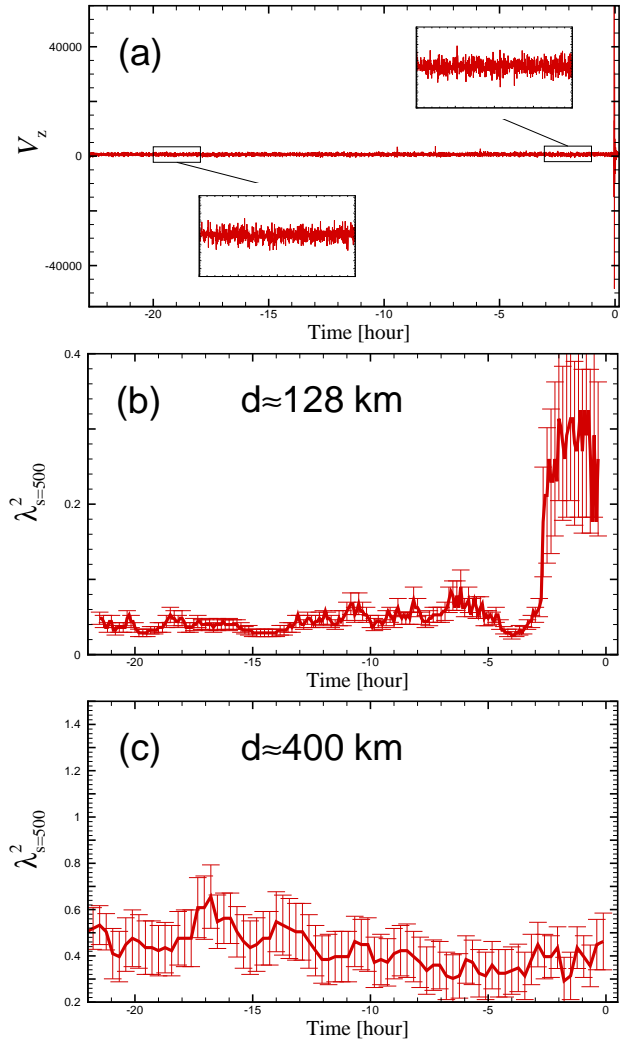}}
\caption{(a) The data for the $M=5.4$ earthquake in California. (b) and (c)
Show the local temporal dependance of $\lambda^2_s$ for $s=500$ ms, collected
at stations with a distance $d$ from the epicenter. The station at $d=$ 128 km
provides the alert, whereas the second station does not yield any alert.}
\end{figure*}

\end{document}